\documentstyle [11pt] {article}

\newcommand{\beq}{\begin{equation}}
\newcommand{\eeq}{\end{equation}}
\newcommand{\beqs}{\begin{eqnarray}}
\newcommand{\eeqs}{\end{eqnarray}}
\newcommand{\prl}{Phys. Rev. Lett.}
\newcommand{\prd}{Phys. Rev. D}

\newcommand{\plb}{Phys. Lett. B}

\begin{document}

\def\thefootnote{\fnsymbol{footnote}}
\baselineskip 6.5mm

\begin{flushright}
\begin{tabular}{l}
ITP-SB-93-63    \\
November, 1993
\end{tabular}
\end{flushright}

\vspace{8mm}
\begin{center}
{\huge \bf  General Determination of Phases in }\\
\vspace{4 mm}
{\huge \bf Leptonic Mass Matrices}

\vspace{4mm}
\vspace{16mm}

\setcounter{footnote}{0}
Alexander Kusenko\footnote{email: sasha@max.physics.sunysb.edu}
\setcounter{footnote}{6}
and Robert Shrock\footnote{email: shrock@max.physics.sunysb.edu}

\vspace{6mm}
Institute for Theoretical Physics  \\
State University of New York       \\
Stony Brook, N. Y. 11794-3840  \\

\vspace{20mm}

{\bf Abstract}
\end{center}

We construct new invariants and give several theorems which determine in
general (i) the number of physically meaningful phases in charged lepton and
neutrino mass matrices and (ii) which elements of these matrices can be
rendered real by rephasings.  We illustrate our results with simple models.

\vspace{35mm}

\pagestyle{empty}
\newpage

\pagestyle{plain}
\pagenumbering{arabic}
\renewcommand{\thefootnote}{\arabic{footnote}}
\setcounter{footnote}{0}

   Understanding fermion masses and mixing remains one of
the most important outstanding problems in particle physics.  In particular,
the issue of possible neutrino masses and associated lepton mixing is of
fundamental interest.  Although there is no definite direct evidence for
nonzero neutrino masses\footnote{For reviews and current limits, see
Refs. \cite{pdg} and \cite{nurev}.  The apparent solar neutrino deficit is the
most suggestive indirect evidence at present. The situation with atmospheric
neutrinos is unclear.} they are expected on general grounds: given only the
known left-handed neutrino fields and the usual Higgs field(s)\footnote{
This comment applies formally to the standard model; however,
the Higgs sector of this model is presumably stabilized by supersymmetry, in
which there are (at least) a pair of Higgs fields with $Y=1$ and $Y=-1$.},
nonzero neutrino masses result generically from
higher-dimension operators which would occur at a scale near to that
of quantum gravity, suppressed by associated inverse powers of the
(reduced) Planck mass,
$\bar M_{P} = \sqrt{\hbar c/(8 \pi G_N)} = 2.4 \times 10^{18}$ GeV,
For example, gauge-invariant dimension-5 operators of this type could
produce neutrino masses of order $m_\nu \sim a v^2/\bar M_P$, where
$v=250$ GeV is the scale of electroweak symmetry breaking, and $a$ is a
dimensionless constant.  Thus, one can understand on general grounds why
neutrino masses are so small.  If electroweak-singlet (EWS) neutrino
fields exist, they would lead, via renormalizable, dimension-4 operators, to
neutrino masses $m_\nu \sim v^2/M_R$, where the scale $M_R$ of the EWS
neutrino mass is naturally $>>v$, again yielding, albeit for a different
reason, very small $m_\nu$ \cite{seesaw}.  In turn, a natural concomitant of
(nondegenerate) neutrino masses is lepton mixing, which is thus
also a generic expectation.\footnote{The lepton mixing angles are functions
of ratios of elements of neutrino matrix elements and of
charged lepton mass matrix elements, and even though left-handed neutrino
masses are small, some of these ratios could, in principle, be $O(1)$.
However, a set of conditions for natural suppression of observable lepton
flavor violation were formulated, and it was shown that the standard model
(generalized to include nonzero $m_\nu$) satisfies these \cite{bwlrs}.}
It is thus of interest to study models for the leptonic mass matrices in the
charge $Q=0$ and $Q=-1$ sectors.  The diagonalization of these mass matrices
determines both the masses and the observable lepton flavor mixing.
In analyzing models of fermion masses and mixing, an important step
is to determine the number of real amplitudes and unremovable, and hence
physically meaningful, phases in the mass matrices.  Surprisingly, there was no
general solution in the literature to the problem of counting the number of
complex phases and determining which elements of mass matrices can be rendered
real by fermion rephasings.  We recently presented a general solution for the
quark sector \cite{ks}.  The leptonic case is qualitatively different and
more complicated because of the general presence of both Majorana and Dirac
masses.  In this Letter we give a solution for the leptonic case.

     The mass terms for the charged leptons can be written in terms of the
$G_{SM}=$ SU(3) $\times$ SU(2) $\times$ U(1) fields as
\beq
-{\cal L}_{mass,Q=-1} = \sum_{j,k=1}^3 \Bigl [\bar L_{2 j L} M_{jk}^{(\ell)}
\ell_{k R}  \Bigr ] + h.c.
\label{ellmass}
\eeq
where $j,k=1,2,3$ denote generation indices,
$L_{j L} = \Bigl (^{\nu_{_{\ell_j}}}_{\ell_j} \Bigr )_L$, $\ell_1=e$,
$\ell_2=\mu$, $\ell_3=\tau$, the first subscript on $L_{a,j L}$ is the
SU(2) index and $M^{(\ell)}$ is the charged lepton mass matrix.\footnote{
Here we use the result from LEP and SLC that there are
three generations of fermions with associated light neutrinos \cite{pdg}.}
In contrast to the charged lepton sector, where one at least knows the
relevant fields, in the neutral lepton sector, one does not.
In addition to the three known left-handed $I=1/2$, $I_3=1/2$ Weyl neutrino
fields $\nu_{jL}$, $j=1-3$, there could be some number $n_s$ of
electroweak-singlet neutrino fields $\chi_{j,R}$, $j=1,2,...n_s$.  The general
neutrino mass matrix is given by
\beq
-{\cal L}_{mass,Q=0} =
 {1 \over 2}(\bar\nu_{_L} \ \bar{\chi^{c}}_{_L})
             \left( \begin{array}{cc}
              M^{(L)} & M^{(D)} \\
              (M^{(D)})^T & M^{(R)} \end{array} \right )\left( \matrix{
\nu^{c}_{_R} \cr \chi_{_R} \cr} \right ) + h.c.
\label{numass}
\eeq
where $\nu_{_L}=( \nu_e,\nu_\mu,\nu_\tau)_L^T$,
$\chi_{_R}=(\chi_1,...,\chi_{n_s})^T_R$; $M_L$ and $M_R$ are $3 \times 3$ and
$n_s \times n_s$ Majorana mass matrices, and $M_D$ is an 3-row by $n_s$-column
Dirac mass matrix.\footnote{In
the context of the (not necessarily minimal) supersymmetric
standard model, we assume unbroken $R$ parity so that the neutrinos do not
mix with the neutralinos (higgsinos and neutral color-singlet gauginos).}
We denote the $(3+n_s) \times (3+n_s)$ neutrino mass
matrix in (\ref{numass}) as $M^{(0)}$.  Thus, $M^{(L)}_{jk}=M^{(0)}_{jk}$,
$j,k=1,2,3$; $M^{(R)}_{jk}=M^{(0)}_{3+j,3+k}$, $j,k=1,...n_s$; and
$M^{(D)}_{jk}=M^{(0)}_{3+j,k}$, $j=1,2,3$, $k=1,..,n_s$, . In general, all of
these matrices are complex. Recall that the anticommutativity of fermion
fields in the path integral and the property $C^T=-C$ (where $C$ is the Dirac
charge conjugation matrix) imply that
\beq
M^{(f)} = M^{(f)T} \ \ , \ \  for \ \ f = L, R
\label{majsym}
\eeq so that
\beq
M^{(0)} = M^{(0)T}
\label{m0sym}
\eeq
The diagonalization of
(\ref{numass}) yields, in general, $3+n_s$ nondegenerate Majorana neutrino
mass eigenstates (some could, of course, be degenerate in magnitude, leading
to possible Dirac neutrino masses).

    Although our theorems on phases are general, some comments are necessary
concerning possible electroweak-singlet neutrinos.  In the supersymmetric
generalization of the standard model, which is strongly motivated by its
ability to maintain the condition $v << M_P$ beyond tree level without
fine-tuning, the component fields $\chi^c_{j,L}$ would arise as part of
$G_{SM}$-singlet chiral superfields $\hat \chi^c_{j}$ (all chiral superfields
will be written as left-handed).  One may recall that, in general, chiral
superfields which are singlets under the standard model gauge group $G_{SM}$
can destabilize the hierarchy \cite{args}.  However, in commonly used models,
the symmetry (such as matter parity) which prevents excessively rapid proton
decay also excludes the types of terms which would destabilize the hierarchy.

    To count the number of physically meaningful phases, we rephase the
lepton fields so as to remove all possible phases in $M^{(\ell)}$ and
$M^{(0)}$ by
\beq
L_{j L} = e^{ - i \alpha_j} L_{j L}'
\label{llrephase}
\eeq
\beq
\ell_{j R} = e^{i \beta^{(\ell)}_j} \ell_{j R}'
\label{lrrephase}
\eeq
for $j=1,2,3$, and, if there exist any $\chi_{j R}$'s,
also
\beq
\chi_{j R} = e^{i \beta^{(\chi)}_j} \chi_{j R}'
\label{chirephase}
\eeq
for $j=1,...n_s$.  In terms of the primed (rephased) lepton fields, the mass
matrices have elements
\beq
M_{jk}^{(\ell) \prime} = e^{i(\alpha_j + \beta^{(\ell)}_k)}M_{jk}^{(\ell)}
\label{mellrephased}
\eeq
for the charged leptons, and, for the neutrino sector,
\beq
M_{jk}^{(L) \prime} = e^{i(\alpha_j + \alpha_k)}M_{jk}^{(L)}
\label{mlrephased}
\eeq
for $j,k=1,2,3$;
\beq
M_{jk}^{(R) \prime} = e^{i(\beta^{(\chi)}_j + \beta^{(\chi)}_k)}M_{jk}^{(R)}
\label{mrrephased}
\eeq
for $j,k=1,...,n_s$, and
\beq
M_{jk}^{(D) \prime} = e^{i(\alpha_j + \beta^{(\chi)}_k)}M_{jk}^{(D)}
\label{mdrephased}
\eeq
for $j=1,2,3$ and $k=1,...,n_s$.

Thus, if $M^{(\ell)}$ has $N_\ell$ nonzero
(and, in general, complex) elements,
then the $N_\ell$ equations for making these elements real are
\beq
\alpha_j + \beta^{(\ell)}_k = -arg(M_{jk}^{(\ell)}) + \eta^{(\ell)}_{jk}\pi
\label{mellrephaseq}
\eeq
where the set $\{jk\}$ runs over each of these nonzero elements,
and $\eta^{(f)}_{jk} = 0$ or $1$.\footnote{The $\eta_{jk}$ term
allows for the possibility of making the
rephased element real and negative; this will not affect the counting of
unremovable phases.}

   Similarly, in the neutrino
sector, if $M^{(D)}$, $M^{(L)}$, and $M^{(R)}$ have, respectively,
$N_D$, $N_L$, and $N_R$ nonzero (and, in general, complex) elements, then the
corresponding equations for making these elements real are
\beq
\alpha_j + \beta^{(\ell)}_k = -arg(M_{jk}^{(D)}) + \eta^{(D)}_{jk}\pi
\label{mdrephaseq}
\eeq
\beq
\alpha_j + \alpha_k = -arg(M_{jk}^{(L)}) + \eta^{(L)}_{jk}\pi
\label{mlrephaseq}
\eeq
and
\beq
\beta^{(\chi)}_j + \beta^{(\chi)}_k = -arg(M_{jk}^{(R)}) + \eta^{(R)}_{jk}\pi
\label{mrrephaseq}
\eeq
where the various ranges of indices are obvious from (\ref{numass}).
Let us define the $(6+n_s)$-dimensional vector of fermion field phases
\beq
v = (\{\alpha_i \}, \{ \beta^{(\ell)}_i \}, \{\beta^{(\chi)}_i\})^T
\label{vvector}
\eeq
where $\{\alpha_i\} \equiv \{\alpha_1, \alpha_2, \alpha_3\}$,
$\{\beta^{(\ell)}_i\} \equiv
\{\beta^{(\ell)}_1, \beta^{(\ell)}_2, \beta^{(\ell)}_3\}$, and
$\{\beta^{(\chi)}_i\} \equiv
\{\beta^{(\chi)}_1,...,\beta^{(\chi)}_{n_s} \}$.
We also define the vector of phases of elements of the various mass matrices
\beq
w=(\{arg(M^{(\ell)})+\eta^{(\ell)}_{jk}\pi \},
\{\{arg(M^{(0)})+\eta^{(0)}_{jk}\pi \})^T
\label{wvector}
\eeq
of dimension equal to the number of rephasing equations $N_{eq} = N_\ell+N_0$,
where $N_0=N_L+N_R+N_D$. We can then write (\ref{mellrephaseq})-
(\ref{mrrephaseq}) as
\beq
T v = w
\label{teq}
\eeq
which defines the $N_{eq}$-row by $(6+n_s)$-column matrix $T$.

   Our first main theorem is:  The number of unremovable and hence physically
meaningful phases $N_p$ in $M^{(\ell)}$ and $M^{(0)}$ is
\beq
N_p = N_{eq}- rank(T)
\label{np}
\eeq
In order for these phases to be physically meaningful, it is, of course,
necessary that they be unchanged under the full set of rephasings of fermion
fields.  In fact, as we will show, there is a one-to-one correspondence between
each such unremovable phase and an independent phase of a certain product of
elements of mass matrices which is invariant under fermion field rephasings.
(In special cases a model may have an unremovable, invariant phase which,
because of a particular symmetry, is zero or $\pi$; we give an example below.)
 The proof of (\ref{np})
is similar to that for the analogous theorem which we proved
for the quark sector \cite{ks}: Let $rank(T)=r_{_T}$.  Then one can delete
$N_{eq}-r_{_T}$ rows from the matrix $T$, i.e. not attempt to remove the
phases from the corresponding elements of the $M^{(\ell)}$ and $M^{(0)}$.
For the remaining $r_{_T}$ equations, one moves a subset of
$6+n_s-(N_{eq}-r_{_T})$ phases in $v$ to the right-hand sides of
(\ref{mellrephaseq})-(\ref{mrrephaseq}), thus including them in a redefined
$\bar w$. This yields a set of $r_{_T}$ linear equations in $r_{_T}$ unknown
phases, denoted $\bar v$. We write this as $\bar T \bar v = \bar w$.  Since,
by construction, $rank(\bar T)=r_{_T}$, $\bar T$ is invertible, so that one can
now solve for the $r_{_T}$ fermion rephasings in $\bar v$ which render
$r_{_T}$ of the $N_{eq}$ complex elements real. Hence there are
$N_{eq}-r_{_T}$ remaining phases in $M^{(\ell)}$ and $M^{(0)}$, as claimed.
$\Box$.\footnote{In the quark case, $T$ is a $N_{eq} \times 9$ matrix, and
we noted \cite{ks} that it could not have maximal rank because one overall
rephasing left the Yukawa (or equivalently, mass) matrices invariant.  In
the present case, unless $M^{(L)}$ and $M^{(R)}$ both vanish, there is no
analogous overall rephasing which leaves $M^{(\ell)}$ and
$M^{(0)}$ invariant, and hence $T$ can have maximal rank.}

   Some comments are in order.  First, as is clear from our proof, in general,
the result  (\ref{np}) does not depend on whether or not
$M^{(f)}_{jk} = M^{(f)}_{kj}$
initially for $f=\ell,D$.  Hence making $M^{(\ell)}$ or $M^{(D)}$ initially
(complex) symmetric does not, in general, result in any reduction in $N_p$,
since in terms of the $G_{SM}$ theory, nothing guarantees this symmetry, and
the rephasing is carried out in terms of $G_{SM}$ fields.  Second, if one of
the unremovable phases is put in a given off-diagonal $M^{(f)}_{pq}$ for
$f=\ell$ or $D$, one may wish to modify the $qp'th$ equation to read
\beq
\alpha_q + \beta^{(f)}_p = -arg(M_{qp}^{(f)})-arg(M_{pq}^{(f)})
\label{mhermitian}
\eeq
for $f=\ell,D$.
For example, in a model where $|M^{(f)}_{pq}|=|M^{(f)}_{qp}|$, this would
yield $M_{pq}^{(f)*} = M_{qp}^{(f)}$ for this pair $pq$ and $f=\ell$ or $D$.
The modification in (\ref{mhermitian}) has no effect on the counting of
phases.  Of course, given the property (\ref{majsym}), when one rephases to
make $M^{(f)}_{jk}$ real for $f=L,R$, this automatically does the same for
$M^{(f)}_{kj}$.

 A fundamental question concerns which elements of $M^{(\ell)}$ and $M^{(0)}$
(taken to be complex initially) can be made real by fermion rephasings.
This is connected with the issue of
which rows are to be removed from $T$ to obtain $\bar T$, i.e.
which nonzero elements of $M^{(\ell)}$ and $M^{(0)}$ are left complex.
We present some more theorems which answer
this question.  The general method is to construct all independent complex
products of elements of $M^{(\ell)}$ and $M^{(0)}$ which are
rephasing-invariant with respect to (\ref{llrephase}), (\ref{lrrephase}) and,
if $\chi_{jR}$'s are present, also (\ref{chirephase}).
These must involve an even number of such elements, since for each index on an
$M$, there must be a corresponding index on an $M^*$ in order to form a
rephasing-invariant product.  Since in general, by construction, these have
arguments $\ne 0,\pi$, each one implies a constraint which is that the
set of $2n$ elements which comprise it cannot be made simultaneously real by
any rephasings or the charged lepton or neutrino fields.  (In special cases, a
symmetry of a model may render some of these real,.) In the neutrino sector,
although the Dirac mass matrix conserves the global lepton number U(1),
while the Majorana mass matrices violate this symmetry,
it is possible and convenient to treat $\nu^c_{j R}$ and $\chi_{k R}$ together,
as was done in (\ref{numass}), the RHS of which can be written simply as
$(1/2)\bar f^{(0)}_L M^{(0)} f^{(0)}_R + h.c.$.  The rephasings
in (\ref{llrephase}) and (\ref{chirephase}) can be
represented in a similarly unified way as $f_{j R} \to e^{i\gamma_j}f_{j R}$.
We thus construct the rephasing invariants
\beq
P^{(\ell)}_{2n;j_1 k_1,...j_n k_n;\sigma_L} =
\prod_{a=1}^{n}M^{(\ell)}_{j_a k_a}M^{(\ell)*}_{\sigma_L(j_a) k_a}
\label{pgeneral}
\eeq
where $f=\ell$ and $\sigma_L$ is an element of the permutation
group $S_n$, and
\beq
P^{(0)}_{2n;j_1 j_2,...,j_{2n-1} j_{2n};\sigma} =
\prod_{a=1}^{n}
M^{(0)}_{j_{2a-1}j_{2a}} M^{(0)*}_{\sigma(j_{2a-1})\sigma(j_{2a})}
\label{p0}
\eeq
where $\sigma \in S_{2n}$.  (The difference in structure is due to
(\ref{m0sym}).)

At quartic order, $2n=4$, (\ref{pgeneral}) and (\ref{p0}) both yield the same
complex invariant, which (in the notation of (\ref{pgeneral})) is
\beq
P^{(f)}_{4;j_1 k_1, j_2 k_2; \tau} \equiv P^{(f)}_{j_1 k_1, j_2 k_2} =
M^{(f)}_{j_1 k_1}M^{(f)}_{j_2 k_2}M^{(f)*}_{j_2 k_1}M^{(f)*}_{j_1 k_2}
\label{p}
\eeq
for $f=\ell,0$.  Note the symmetries
\beq
P^{(f)}_{j_1 k_1, j_2 k_2}=P^{(f)}_{j_2 k_2, j_1 k_1}
\label{pswitch}
\eeq
\beq
P^{(f)}_{j_1 k_1 j_2 k_2}=P^{(f)*}_{j_1 k_2, j_2 k_1}
\label{pconjugate}
\eeq
(whence also $P^{(f)}_{j_1 k_1, j_2 k_2} = P^{(f)*}_{j_2 k_1, j_1 k_2}$) for
$f=\ell, 0$ and, for $f=0$, the additional symmetry
\beq
P^{(0)}_{j_1 j_2, j_3 j_4} = P^{(0)}_{j_2 j_1, j_4 j_3}
\label{p0sym}
\eeq
In a shorthand notation, (\ref{p}) for $f=\ell$ may be denoted $\ell
\ell \ell^* \ell^*$.
It is useful to classify the complex $P^{(0)}$ invariants according to which
submatrices $-$ $M^{(L)}$, $M^{(R)}$, and/or $M^{(D)}$ $-$ in $M^{(0)}$
they involve. We find five possible (complex) structures:
$L L L^* L^*$, $R R R^* R^*$, $L D L^* D^*$, $R D R^* D^*$, and $L R D^* D^*$.
Explicitly, the first two are given by (\ref{p}) for $f=L,R$, respectively, and
the others by the products
\beq
\Pi^{(L f)}_{j_1 j_2, j_3 m_1} =
M^{(L)}_{j_1 j_2}M^{(f)}_{j_3 m_1}M^{(L)*}_{j_1 j_3}M^{(fnn)*}_{j_2 m_1}
\label{rlf}
\eeq
for $f=D$,
\beq
\Xi^{(R D)}_{k_1 k_2, j_1 k_3} =
M^{(R)}_{k_1 k_2}M^{(D)}_{j_1 k_3}M^{(R)*}_{k_1 k_3}M^{(D)*}_{j_1 k_2}
\label{srd}
\eeq
and
\beq
\Omega^{(LRDD)}_{j_1 j_2, k_1 k_2} = M^{(L)}_{j_1 j_2}M^{(R)}_{k_1 k_2}
M^{(D)*}_{j_1 k_1}M^{(D)*}_{j_2 k_2}
\label{tlrdd}
\eeq
In terms of $P^{(0)}$'s (and using (\ref{m0sym}), these are
\beq
\Pi^{(L D)}_{j_1 j_2, j_3 m_1} = M^{(0)}_{j_1,j_2}M^{(0)}_{j_3,3+m_1}
M^{(0)*}_{j_1,j_3}M^{(0)*}_{j_2,3+m_1} = P^{(0)}_{j_2 \ j_1, \ j_3 \ 3+m_1}
\label{pirel}
\eeq
\beq
\Xi^{(R D)}_{k_1 k_2, j_1 k_3} = M^{(0)}_{3+k_1,3+k_2}M^{(0)}_{j_1,3+k_3}
M^{(0)*}_{3+k_1,3+k_3}M^{(0)*}_{j_1,3+k_2} =
P^{(0)}_{3+k_1 \ 3+k_2, \ j_1 \ 3+k_3}
\label{xirel}
\eeq
\beq
\Omega^{(LRDD)}_{j_1 j_2, k_1 k_2} =  M^{(0)}_{j_1,j_2}
M^{(0)}_{3+k_1,3+k_2}M^{(0)*}_{j_1,3+k_1}M^{(0)*}_{j_2,3+k_2} =
P^{(0)}_{j_2 \ j_1, \ 3+k_1 \ 3+k_2}
\label{omrel}
\eeq

Secondly, we construct a set of complex invariants connecting the $Q=-1$ and
$Q=0$ lepton sectors.  At order $2n=4$ we find two independent types of complex
invariants, which we denote as
\beq
Q^{(D \ell)}_{j_1 k_1,j_2 m_1} =
M^{(D)}_{j_1 k_1}M^{(\ell)}_{j_2 m_1}M^{(D)*}_{j_2 k_1}M^{(\ell)*}_{j_1 m_1}
\label{qdll}
\eeq
and $\Pi^{L \ell}_{j_1 j_2, j_3 m_1}$ as given by (\ref{rlf}) for $f=\ell$.
Note that
\beq
Q^{(D \ell)}_{j_1 k_1, j_2 m_1}=Q^{(D \ell)*}_{j_2 k_1, j_1 m_1}
\label{qdllsym}
\eeq
In the short notation, these are $D \ell D^* \ell^*$ and $L \ell L^* \ell^*$.
We also observe that
\beq
\Pi^{(L f)}_{j_1 j_2, j_3 m_1} = Q^{(L,f)}_{j_2 j_1, j_3 m_1}
\label{rq}
\eeq

    At order $2n=6$, we find first the complex invariants
$f f f f^* f^* f^*$ for $f=\ell,0$, each of which involves only one of the two
charge sectors.  The various possibilities for the permutations in
(\ref{pgeneral}) and (\ref{p0}) yield only one independent explicit form, which
we denote simply by
\beq
P^{(f)}_{j_1 k_1, j_2 k_2, j_3 k_3} =
M^{(f)}_{j_1 k_1}M^{(f)}_{j_2 k_2}M^{(f)}_{j_3 k_3}
M^{(f)*}_{j_2 k_1}M^{(f)*}_{j_3 k_2}M^{(f)*}_{j_1 k_3}
\label{p6}
\eeq
The $f=\ell$ invariant is denoted $\ell \ell \ell \ell^* \ell^* \ell^*$. The
$f=0$ invariants can be further characterized according to which submatrices of
$M^{(0)}$ they involve; we find the following types:
$LLLL^*L^*L^*$,
$RRRR^*R^*R^*$,
$DDDD^*D^*D^*$,
$D D L D^* D^* L^*$,
$D D R D^* D^* R^* $,
$L L D L^* L^* D^*$,
$R R D R^* R^* D^* $,  \newline
$L R D L^* R^* D^*$,
$L L R L^* D^* D^*$,
$L R R D^* D^* R^*$, and
$L R D D^* D^* D^*$.
(In our notation, $fff' f^*f^*f^{\prime *}$ and $f'fff^{\prime *}f^*f^*$ are
synonymous; also, $L D D L^* L^* R^*$ contains the same phase
information as $L L R L^* D^* D^*$, and so forth for the other conjugates of
this type, which are hence not listed.)
For any $f$, the invariants in (\ref{p6}) satisfy
\beq
P^{(f)}_{j_1 k_1, j_2 k_2, j_3 k_3} = P^{(f)}_{j_2 k_2, j_3 k_3, j_1 k_1} =
P^{(f)}_{j_3 k_3, j_1 k_1, j_2 k_2}
\label{p6cycl}
\eeq
and
\beq
P^{(f)}_{j_1 k_1,j_2 k_2, j_3 k_3} = P^{(f)*}_{j_3 k_2, j_2 k_1, j_1 k_3}
\label{p6sym}
\eeq
For the cases where $M^{(f)}$ is symmetric, i.e. $M^{(f)}=M^{(0)}$ or, for
submatrices, $M^{(L)}, M^{(R)}$, we find the additional symmetry
\beq
P^{(f)}_{j_1 j_2, j_3 j_4, j_5 j_6}=P^{(f)}_{j_2 j_1, j_6 j_5, j_4 j_3}
\label{p6maj}
\eeq

We also find the following independent 6'th order complex invariants linking
the charge $Q=-1$ and $Q=0$ sectors:
$\ell \ell L \ell^* \ell^* L^*$,
$L L \ell L^* L^* \ell^* $,
$\ell \ell D \ell^* \ell^* D^*$,
$D D \ell D^* D^* \ell^*$,
$\ell D L \ell^* D^* L^* $,
$\ell D R \ell^* D^* R^* $,
$\ell L R \ell^* D^* D^*$.
The 6'th order mixed-charge invariants of the form $f f f' f^* f^* f^{*'}$,
and also the 6'th order $Q=0$ invariants which are of this form when written in
terms of $L$, $R$ and/or $D$ submatrices, are given explicitly as
\beq
Q^{(fff')}_{j_1 k_1, j_2 k_2, j_3 m_1} =
M^{(f)}_{j_1 k_1}M^{(f)}_{j_2 k_2}M^{(f')}_{j_3 m_1}
M^{(f)*}_{j_1 k_2}M^{(f)*}_{j_3 k_1}M^{(f')*}_{j_2 m_1}
\label{q6}
\eeq
Some examples of explicit forms of other invariants include
\beq
\ell D R \ell^* D^* R^*: \
M^{(\ell)}_{j_1 k_1}M^{(D)}_{j_2 m_1}M^{(R)}_{m_2 m_3}
M^{(\ell)*}_{j_2 k_1}M^{(D)*}_{j_1 m_2}M^{(R)*}_{m_1 m_3}
\label{elr}
\eeq
\beq
\ell L R \ell^* D^* D^*: \
M^{(\ell)}_{j_1 k_1}M^{(L)}_{j_2 j_3}M^{(R)}_{m_1 m_2}
M^{(\ell)*}_{j_2 k_1}M^{(D)*}_{j_1 m_1}M^{(D)*}_{j_3 m_2}
\label{elredd}
\eeq
\begin{eqnarray}
\ell D L \ell^* D^* L^*: \ &
M^{(\ell)}_{j_1 k_1}M^{(D)}_{j_2 m_1}M^{(L)}_{j_3 j_4}
M^{(\ell)*}_{j_2 k_1}M^{(D)*}_{j_3 m_1}M^{(L)*}_{j_1 j_4}
\label{edl}  & \\
& M^{(\ell)}_{j_1 k_1}M^{(D)}_{j_2 m_1}M^{(L)}_{j_3 j_4}
M^{(\ell)*}_{j_3 k_1}M^{(D)*}_{j_4 m_1}M^{(L)*}_{j_1 j_2},\ &  etc. \nonumber
\end{eqnarray}
Further discussion of higher order invariants is presented elsewhere
\cite{lphaselong}.

   Our theorems are as follows:
For a given model, construct the maximal set of independent generally complex
invariants; denote the number of these as $N_{inv}$. Of the corresponding
$N_{inv}$ arguments, a subset $N_{ia}$ are linearly independent. Then
(a) the number of unremovable, invariant, and hence physically meaningful
phases  $N_p$ is equal to $N_{ia}$; (b) in the general case in which the
$N_{ia}$ arguments are $\ne 0, \pi$, each implies a constraint that elements
contained within the corresponding invariant cannot all be made
simultaneously real; (b) this constitutes the complete set of
constraints on which elements of $M^{(\ell)}$ and $M^{(0)}$ can be made
simultaneously real.  Often, the lowest-order nonvanishing complex invariants
are quartic, but we give an exception below.

   In order to determine $N_{ia}$ from the $N_{inv}$ independent complex
invariants, we proceed as follows.  Form the $N_{eq}$ vector of arguments of
nonzero (and in general complex) elements of $M^{(\ell)}$ and $M^{(0)}$, $\xi$.
For each complex invariant of a given order, $X$, $arg(X) =
\sum_{f}\sum_{j,k}c^{(f)}_{j,k} arg(M^{(f})_{jk}$, where the sum is over the
$N_{eq}$ nonzero elements of $M^{(\ell)}$ and $M^{(0)}$. These equations
can be written as $Z \xi = w$, where $\xi$ is the $N_{inv}$-dimensional
vector $\xi = (arg(X_1),...,arg(X_{N_{inv}}))$ and
$Z$ is an $N_{inv}$-row by $N_{eq}$-column matrix.  Then $rank(Z)=N_{ia}$.

    We illustrate our general theorems with some models.\footnote{Since our
purpose is to illustrate our general theorems, we do not attempt to discuss an
exhaustive set of viable models of neutrino masses and mixing.  Indeed, given
the unsettled status of atmospheric neutrino oscillations (claimed by some
experiments, not by others), it is unclear what data a viable model must fit.
Here we shall tentatively
assume that the solar neutrino deficit does indicate neutrino
oscillations but shall view it as premature to make any inference of new
physics from the atmospheric neutrino oscillation data.}  We first consider
the case $n_s=0$, i.e. no electroweak-singlet neutrinos.  For models of this
type, $v$ is 6-dimensional, and $T$ is $N_{eq} \times 6$ dimensional.
As an illustration, consider a model defined by
\beq
M^{(\ell)} =  \left (\begin{array}{ccc}
                  0 & E_{12} & 0 \\
                  E_{21} & E_{22} & E_{23} \\
                  0 &  E_{32} &  E_{33} \end{array}   \right  )
\label{me2}
\eeq
\beq
M^{(L)} =  \left (\begin{array}{ccc}
                  0 & L_{12} & 0 \\
                  L_{12} & L_{22} & L_{23} \\
                  0 &  L_{23} &  L_{33} \end{array}   \right  )
\label{ml2}
\eeq
(We have verified that this model is able to fit current solar neutrino data
and other established limits on neutrino masses and mixing.)
For this model, there are $N_{eq}=10$ complex elements in $M^{(\ell)}$ and
$M^{(D)}$ and corresponding rephasing equations.  The matrix $T$ is $10 \times
6$, and we find that it has rank 6.  Our theorem (\ref{np}) then implies that
there are $N_p=N_{eq}-rank(T)=4$ unremovable phases in $M^{(\ell)}$ and
$M^{(L)}$.  We find $N_{inv}=8$ independent complex (quartic) invariants:
$P^{(\ell)}_{22,33}=E_{22}E_{33}E_{32}^*E_{23}^*$,
$P^{(0)}_{22,33}=P^{(L)}_{22,33}=L_{22}L_{33}L_{23}^{*2}$,
$\Pi^{(L \ell)}_{12,22}=L_{12}E_{22}L_{22}^*E_{12}^*$,
$\Pi^{(L \ell)}_{12,32}=L_{12}E_{32}L_{23}^*E_{12}^*$
$\Pi^{(L \ell)}_{22,32}=L_{22}E_{32}L_{23}^*E_{22}^*$
$\Pi^{(L \ell)}_{22,33}=L_{22}E_{33}L_{23}^*E_{23}^*$,
$\Pi^{(L \ell)}_{23,32}=L_{23}E_{32}L_{33}^*E_{22}^*$,
$\Pi^{(L \ell)}_{23,33}=L_{23}E_{33}L_{33}^*E_{23}^*$. From these we
calculate the $8 \times 10$ dimensional $Z$ matrix and find
that it has rank 4, so that of the eight arguments of the complex invariants,
there are $N_{ia}=4$ linearly independent ones, in accord with the equality
$N_{ia}=N_p$ and our result that $N_p=4$. Thus, for this model, it is not,
in general, possible by any rephasings to make either the charged lepton or
neutrino mass matrices real.  Furthermore, the last complex invariant implies
that it is not possible in general to make the set of elements
$\{L_{23},L_{33},E_{23},E_{33}\}$ simultaneously real; the next-to-last implies
the same for the set $\{L_{23},L_{33},E_{22},E_{32}\}$, and so forth for each
of the other complex invariants.

   A natural setting for models with electroweak-singlet neutrinos is an
SO(10)grand unified theory (GUT), where a $\chi^c_{j L}$ occurs, along with the
SU(5) ${\bf 10}_L$ and ${\bf \bar 5}_L$ in the SO(10) ${\bf 16}_L$, for
each of the three generations, so that $n_s=3$.  An interesting model of this
type was recently studied in a supersymmetric context in Ref. \cite{dhr}
(DHR).  A pioneering study of a non-supersymmetric SO(10) model which was
similar but had more mass matrix parameters was carried out in Ref.
\cite{r}.  In both cases the Higgs representations were restricted so as to
obtain (complex) symmetric mass matrices.  Here we shall consider two models.
First, to exhibit the generality of our results, we will consider a
generalization of the models of Refs. \cite{dhr} and {r} in which $M^{(\ell)}$
and $M^{(D)}$ are not assumed to be symmetric (recall from (\ref{majsym})
that $M^{(R)}$ is automatically symmetric)
\beq
M^{(\ell)} =  \left (\begin{array}{ccc}
                  0 & E_{12} & 0 \\
                  E_{21} & E_{22} & 0 \\
                  0 &  0 &  E_{33} \end{array}   \right  )
\label{memod}
\eeq
\beq
M^{(D)} =  \left (\begin{array}{ccc}
                  0 & D_{12} & 0 \\
                  D_{21} & 0 & D_{23} \\
                  0 &  D_{32} &  D_{33} \end{array}   \right  )
\label{mlmod}
\eeq
\beq
M^{(R)} =  \left (\begin{array}{ccc}
                  0 & R_{12} & 0 \\
                  R_{12} & 0 & 0 \\
                  0 &  0 &  R_{33} \end{array}   \right  )
\label{mdmod}
\eeq
(Here all entries are in general complex.)  Second, we shall consider the DHR
model.  We note that, to reduce the number of parameters, the Yukawa
couplings were further restricted in the DHR model so that
(besides the conditions $M^{(\ell)}=M^{(\ell)T}$ and $M^{(D)}=M^{(D)T}$)
$D_{12} = -3R_{12}$ and $D_{33} = -3R_{33}$.
For the general model (\ref{memod})-(\ref{mdmod}) we calculate the
$11 \times 9$ dimensional $T$ matrix and find that it has rank 9, so that by
our theorem (\ref{np}) there are $N_p=N_{eq}-rank(T)=2$ unremovable, physically
meaningful phases in the leptonic sector. Correspondingly, we find two
independent complex invariants (which have independent arguments), so
$N_{inv}=N_{ia}=2$.  These are both of 6'th order:
$P^{(0)}_{24,53,66} = D_{21}D_{32}R_{33}D_{23}^*D_{33}^*R_{12}^*$ and
$Q^{(DD\ell)}_{32,23,12}=D_{32}D_{23}E_{12}D_{12}^*D_{33}^*E_{22}^*$.
As can be checked explicitly here, making $M^{(\ell)}$ and $M^{(D)}$ (complex)
symmetric does not change the value of $N_p$ or make either of the complex
invariants real.  For the general model (and for the special case in which
$M^{(\ell)}$ and $M^{(D)}$ are symmetric), one thus has the constraints that
the elements in the set
$\{D_{21},D_{23},D_{32},D_{33},R_{12},R_{33}\}$ cannot be made
simultaneously real by any fermion rephasing, and similarly, the elements in
the set $\{D_{12},D_{23},D_{32},D_{33},E_{12},E_{22}\}$ cannot be made
simultaneously real by any rephasings; the first constraint shows that it is
not possible to make both the Dirac neutrino mass matrix $M^{(D)}$ and the
right-handed Majorana mass matrix $M^{(R)}$ simultaneously real by any
fermion field rephasings. For the light particle sector of the general
model, our constraints allow one to place (i) both phases in
the light neutrino mass matrix $M_\nu = M^{(D)}M^{(R)-1}M^{(D)T}$ (with
$M^{(\ell)}$ being made real); or (ii) one phase in $M_{\nu}$ and one in
$M^{(\ell)}$; but (iii) not both in $M^{(\ell)}$.  In contrast, for the DHR
model, it was noted \cite{dhr} that there is only one (complex) phase in the
leptonic sector (and indeed in the model as a whole).  In terms of our
formalism, this follows because, as a result of the relation between $M^{(D)}$
and $M^{(R)}$ given above, $P^{(0)}_{24,53,66}$ is actually real, so that its
phase, while still an unremovable, invariant, and hence meaningful phase,
is no longer complex.  It is still technically true that $N_p=N_{ia}=2$ in this
model, but only one of these phases is complex.
This model (in both its general form and the above specializations) also
serves as an example of one in which there are no complex quartic invariants.
We have shown that this is true of the quark sector of the model, where
we found \cite{ks} one 6'th order invariant, which, written in terms of
Yukawa matrix elements, is $Q^{(uud)}_{32,23,12} =
Y^{(u)}_{32}Y^{(u)}_{23}Y^{(d)}_{12}Y^{(u)*}_{12}Y^{(u)*}_{33}
Y^{(d)*}_{22}$.\footnote{Note that in the models of\cite{dhr} and \cite{r},
$arg(Q^{(uud)}_{32,23,12})=arg(Q^{(DD\ell)}_{32,23,12})-\pi$ so that this
does not represent a new phase in the theory as a whole.}
Further results and applications will be reported elsewhere
\cite{lphaselong}.

   This research was supported in part by NSF grant PHY-93-09888.

\end{document}